\documentclass[twocolumn,secnumarabic,amssymb,nobibnotes,aps,prd,showpacs]{revtex4-1}

\usepackage{graphicx}
\usepackage{amsmath}
\usepackage{float}
\usepackage{longtable}
\newcommand{\be}{\begin{equation}}
\newcommand{\ee}{\end{equation}}
\newcommand{\bea}{\begin{eqnarray}}
\newcommand{\eea}{\end{eqnarray}}

\newcommand{\nn}{\\ \nonumber}
\newcommand{\non}{\nonumber}

\newcommand{\bdtm}{b^{\dagger}_{2\mu}}

\newcommand{\bdt}{b^{\dagger}_{2}}
\newcommand{\bt}{b_{2}}
\newcommand{\lb}{\left(}
\newcommand{\rb}{\right)}
\newcommand{\lbb}{\left[}
\newcommand{\rbb}{\right]}
\newcommand{\pjg}{\varphi_{J}^{\lb g \rb}}
\newcommand{\pjgo}{\varphi_{J_{1}}^{\lb g \rb}}
\newcommand{\pjgt}{\varphi_{J_{2}}^{\lb g \rb}}

\newcommand{\njg}{\mathcal{N}_{J}^{\lb g \rb}}
\newcommand{\njgo}{\mathcal{N}_{J_{1}}^{\lb g \rb}}
\newcommand{\njgt}{\mathcal{N}_{J_{2}}^{\lb g \rb}}

\newcommand{\ijz}{\mathcal{I}_{J}^{\lb 0 \rb}\lb d \rb}
\newcommand{\ijo}{\mathcal{I}_{J}^{\lb 1 \rb}\lb d \rb}

\begin{document}

\title{Description of electromagnetic and favored $\alpha$-transitions in heavy odd-mass nuclei}

\author{A. Dumitrescu$^{1,4}$ and D.S. Delion$^{1,2,3}$}
\affiliation{
$^1$ "Horia Hulubei" National Institute of Physics and Nuclear 
Engineering, \\ 
407 Atomi\c stilor, POB MG-6, Bucharest-M\u agurele, RO-077125, Rom\^ania \\
$^2$ Academy of Romanian Scientists,
54 Splaiul Independen\c tei, Bucharest, RO-050094, Rom\^ania \\
$^3$ Department of Biophysics, Bioterra University, 81 G\^arlei str., Bucharest, RO-013724, Rom\^ania \\
$^4$ Department of Physics, University of Bucharest,
405 Atomi\c stilor, POB MG-11, Bucharest-M\u agurele, RO-077125, Rom\^ania }

\begin{abstract}
We describe electromagnetic and favored $\alpha$-transitions to rotational bands in odd-mass nuclei
built upon a single particle state with angular momentum projection $\Omega\ne \frac{1}{2}$ in the region $88 \le Z \le 98$. 
We use the particle coupled to an even-even core approach described by the Coherent State Model (CSM) and
the coupled channels method to estimate partial $\alpha$-decay widths. 
We reproduce the energy levels of the rotational band where favored $\alpha$-transitions
occur for 26 nuclei and predict $B\lb E2 \rb$ values for electromagnetic transitions
to the bandhead using a deformation parameter and a Hamiltonian strength parameter for each nucleus, together with 
an effective collective charge depending linearly on the deformation parameter. Where
experimental data is available, the contribution of the single particle effective charge to the total $B \lb E2 \rb$ value
is calculated. The Hamiltonian describing the $\alpha$-nucleus interaction contains two terms, a spherically symmetric
potential given by the double-folding of the M3Y nucleon-nucleon interaction plus a repulsive core
simulating the Pauli principle and a quadrupole-quadrupole (QQ) interaction. The $\alpha$-decaying state
is identified as a narrow outgoing resonance in this potential. The intensity of the transition to the first excited state 
is reproduced by the QQ coupling strength. It depends linearly both on the nuclear deformation
and the square of the reduced width for the decay to the bandhead, respectively. 
Predicted intensities for transitions to higher excited states are in a reasonable agreement with experimental data. 
This formalism offers a unified description of energy levels, electromagnetic and favored $\alpha$-transitions for 
known heavy odd-mass $\alpha$-emitters.
\end{abstract}


\maketitle

\section{Introduction}

A brief overview of the $\alpha$-emission process in even-even nuclei is helpful for the understanding of the 
more complex situation in odd-mass emitters. 

In the case of transitions to excited states, the single-particle levels around 
the Fermi surface play the dominant role and the corresponding decay widths are very sensitive
to the structure of the daughter nucleus \cite{Del10,Buc12}.
An important problem is the study of the $\alpha$-daughter interaction. One of the most popular approaches is
the double folding procedure \cite{Neu92}. This method has been used together with the coupled channels approach and a
repulsive core simulating the Pauli principle in order to study the $\alpha$-decay fine structure 
in transitional and rotational even-even nuclei \cite{Del06}. 
For a thorough study of the structure and $\alpha$-emission spectrum in vibrational, transitional
and rotational even-even nuclei, see Ref. \cite{Del15}.

Several calculations for the fine structure of the emission spectrum have already been made in the case of
odd-mass $\alpha$-emitters. For example, in Ref. \cite{Ni12}
a multichannel cluster model together with the coupled channels equation is used to calculate branching ratios to excited states
for favored transitions in heavy emitters, in the region $93<Z<102$.
In Ref. \cite{War15}, a microscopic method is employed with a Skyrme SLy4 effective interaction. Starting from the Hartree-Fock-Bogoliubov
vacuum and quasiparticle excitations, the $\alpha$-particle formation amplitude is calculated for the $\alpha$-decay to various
channels mostly in the $84<Z<88$ region. Several unfavored transitions are treated in this paper and predictions are made for the
properties of the g.s.$\rightarrow$g.s. $\alpha$-trasition in odd-mass superheavy nuclei.
The unfavored g.s.$\rightarrow$g.s. $\alpha$-decay in odd-mass nuclei in the region $64\le Z\le 112$ is also treated in Ref. \cite{Sei15}, 
with the purpose of investigating the effect of the difference in the spin and parity of the ground states on the $\alpha$-particle
and daughter nucleus preformation probability. The calculations are done in the framework of the extended cluster model,
with the Wentzel-Kramers-Brillouin penetrability and assault frequency, together with an interaction potential computed on the basis
of the Skyrme SLy4 interaction.

In the current paper, we expand the method previously used in Ref. \cite{Del15} for the even-even case by
allowing the coupling of an odd-particle to a core described by a coherent function. 
We study the energy levels and electromagnetic transition
rates of this nucleus and then couple an $\alpha$-particle to it in order to describe
the emission spectrum for the case of favored transitions.
Our method is to employ an $\lb I,l \rb$ coupling procedure in the laboratory frame, between the
angular momentum $I$ of the daughter nucleus and the orbital angular momentum $l$ of the $\alpha$-particle, similar to
Nilsson's original $\lb I, j \rb$ coupling method for the description of nuclear spectra in the intrinsic frame \cite{Nil}, where
$j$ is the angular momentum of the odd particle. We show that using a small basis having a single value for
$l$ in each channel, we can use a QQ $\alpha$-daughter interaction to
generate simultaneously resonant states of even or odd parity at the same reaction energy and QQ coupling strength.
The partial decay widths obtained this way are in good agreement with the available experimental data.

\section{Theoretical Background}

In this section we present the theoretical tools required for the calculation
of energy levels and electromagnetic transition rates for odd-mass nuclei,
as well as the coupled-channels method that is applied to the study of the 
fine structure of the $\alpha$-emission spectrum.

\subsection{Nucleon coupled to a coherent state core}

A description of the surface dynamics of a deformed even-even nucleus was
proposed for the first time in Refs. \cite{Lip69,Lip76} by using a coherent 
state of quadrupole bosons. A generalization to
all types of low-energy collective motion was proposed in Refs. \cite{Rad76a,Rad76b}
and was extensively developed in Refs. \cite{Rad81,Rad82} as the coherent state
model (CSM). A review paper on this topic is available in Ref. \cite{Rad12}, as well
as in the textbook \cite{Rad14}. Here, we will present in a concise manner the main
ideas of the model, and then extend them to the coupling of an additional nucleon 
to the even-even core. The final goal is to describe a rotational band built upon
a given single-particle state of the odd nucleon.

We begin by assuming that the intrinsic state of an axially deformed
even-even nucleus is given by a coherent superposition of quadrupole bosons 
$\bdtm$ with $\mu=0$

\be\label{incst}
|\phi_{g}\rangle=e^{d\lb b^{\dag}_{20}-b_{20} \rb }|0\rangle,
\ee
where $|0\rangle$ is the phonon vacuum and the quantity $d$ is
called deformation parameter \cite{Rad81}.

The physical states that define the ground band are obtained by angular momentum
projection

\be
|\varphi_{JM}^{\lb g \rb}\rangle=\mathcal{N}_{J}^{\lb g \rb}P^{J}_{M0}|\phi_{g}\rangle.
\ee

$P_{M0}^{J}$ is the projection operator which has the integral representation

\be
P^{J}_{MK}=\sqrt{\frac{2J+1}{8\pi^{2}}}\int {d\omega}\mathcal{D}^{J}_{MK}\lb \omega \rb R \lb \omega \rb,
\ee
with $\omega$ the set of three Euler angles, $\mathcal{D}^{J}_{MK}\lb \omega \rb$ a Wigner function
and $R \lb \omega \rb$ the rotation operator.

$\njg$ is the norm of the projected state, given by the formula

\be
\njg=\lbb \hat{J}^{2} \ijz  \rbb^{-\frac{1}{2}}e^{\frac{d^{2}}{2}},
\ee
with $\hat{J}=\sqrt{2J+1}$ and $\ijz$ given by

\be
\ijz=\int_{0}^{1}\mathcal{P}_{J}\lb x \rb e^{d^{2}\mathcal{P}_{2}\lb x \rb}dx,
\ee
in terms of the Legendre polynomial $\mathcal{P}_{J}$.

For an odd-mass nucleus, the state of total angular momentum $I$ and
projection $M$ is given by projecting out the product between
the coherent state (\ref{incst}) and the single particle state
$\psi_{jm}$, where $j$ is a shorthand notation for all of the quantum numbers
of the state, that is

\be
\Phi_{IM}=P^{I}_{M0}\lbb \psi_{j} \phi_{g} \rbb.
\ee

A straighforward calculation leads to the following result
\be\label{phim}
\Phi_{IM}=\sum_{J}X_{I}^{Jj}\lbb \pjg \otimes \psi_{jm} \rbb_{IM},
\ee
with normalization coefficients $X_{I}^{Jj}$ given by

\be
X_{I}^{Jj}=
\frac{\lb \njg \rb^{-1}\langle jJ;\Omega 0|I \Omega \rangle}
{\sqrt{\sum\limits_{J'}\lb \mathcal{N}_{J'}^{\lb g \rb}\rb ^{-2} \lb \langle j J' \Omega 0 | I \Omega  \rangle \rb^{2}}},
\ee
where the bra-ket product denotes a Clebsch-Gordan coefficient and $\Omega$
is the fixed z-projection of the single-particle angular momentum $j$. More details on this procedure
can be consulted in Ref. \cite{Del93}.

The states built upon the bandhead $I=j=\Omega$ that follow the sequence
$I=\Omega,\Omega+1,\Omega+2,\dots$ constitute a rotational band. In the
Nilsson model, these states are labeled by the set $\Omega^{\pi}\lbb N n_{z} \Lambda \rbb$,
where $\pi$ is the parity, $N$ is the principal quantum number, $n_{z}$ the number of
nodes of the radial wavefunction in the $z$ direction and $\Lambda$ the projection of the
single-particle orbital angular momentum. The last three numbers act only as labels, as the
good quantum numbers are only $\Omega$ and $\pi$. 

The simplest Hamiltonian that can describe such a rotational structure consists of two terms \cite{Del93}:

\be\label{ham}
H=A_{1}\bdt\cdot\bt-A_2r^2\lb \bdt+\tilde{b}_{2}\rb\cdot Y_{2}.
\ee
where by dot we denoted the scalar product.
$A_{1}$ is a strength parameter required to fit experimental data and $A_2$ is the
strength of the particle-core QQ interaction. For a given ladder operator
$a_{l}$, we have

\be
\tilde{a}_{l\mu}=\lb - \rb^{\mu}a_{l -\mu}.
\ee

For the description of the rotational band the only relevant parameter is $A_1$
due to the fact that the particle-core term is common. Instead of solving the eigenvalue problem by a full diagonalization procedure,
a simpler approach, involving the analytical expression for the diagonal matrix elements of the Hamiltonian
(\ref{ham}) in the basis of Eq. (\ref{phim}) suffices:

\bea
\label{HH}
\langle IM|H|IM\rangle&=& A_{1}d^{2}f_{j\Omega I}- d \lb N+ \frac{3}{2} \rb\times\\
\non&\times& \langle j 2; \Omega 0| j \Omega\rangle \langle j 2; \frac{1}{2} 0| j \frac{1}{2} \rangle,
\eea
with $f_{j\Omega I}$ given by

\be
f_{j \Omega I}=\frac{\sum\limits_{J}\langle I j; \Omega -\Omega | J 0 \rangle^{2}\ijo}
{\sum\limits_{J}\langle I j; \Omega -\Omega | J 0 \rangle^{2} \ijz },
\ee
in terms of the function

\be
\mathcal{I}_{J}^{\lb 1 \rb}\lb x \rb=\frac{d}{dx}\mathcal{I}_{J}^{\lb 0 \rb}\lb x \rb.
\ee

\begin{figure} 
\begin{center} 
\includegraphics[width=9cm]{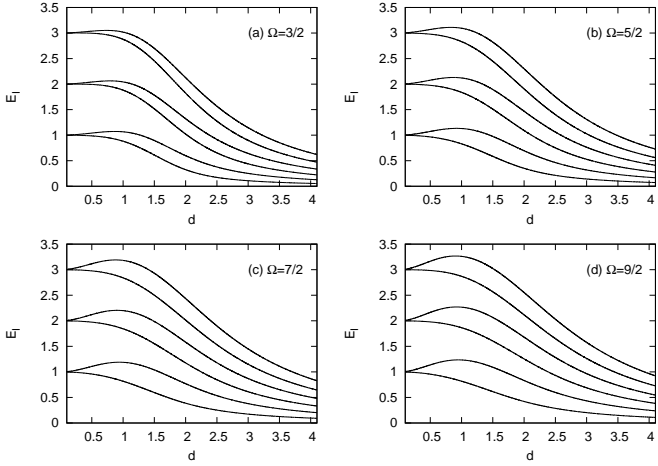}
\vspace{-8mm}
\caption{Normalized energy levels $E_{I}$ as function of deformation $d$, for different values of the single particle angular
momentum projection $\Omega$.}
\label{fig1}
\end{center} 
\end{figure}

The shape of such a spectrum is dependent both on the deformation parameter and on the value of $\Omega$, as can be seen in Fig. \ref{fig1}.

While this approach is adequate for the purpose of this paper, if a greater
precision in the description of the nuclear energy spectrum is required, then more
terms can be added to the Hamiltonian (\ref{ham}), as shown in Ref. \cite{Del93}. 
Let us also mention that the development presented here and expanded upon in Ref. \cite{Del93} is appropriate for any rotational band built 
upon an angular momentum projection $\Omega \ne \frac{1}{2}$. The special case $\Omega=\frac{1}{2}$ 
requires a modification of the formalism and will be treated in a future paper.

\subsection{Electromagnetic transitions}

The $B \lb E2 \rb$ values of electric quadrupole transitions follow from both
collective and single particle contributions

\bea
B\lb E2; I_{2}\rightarrow I_{1} \rb&=&
[ \frac{1}{\hat{I}_{2}} \langle I_{1} || q_{0}^{c}Q^{c}_{2} || I_{2} \rangle  +\\
\non &+& \frac{1}{\hat{I}_{2}} \langle I_{1} || q_{0}^{sp}Q^{sp}_{2} || I_{2} \rangle  ]^{2},
\eea
where $q_{0}^{c}$ and $q_{0}^{sp}$ are effective charges.

The collective quadrupole transition operator has both harmonic and anharmonic contributions
\be
Q^{c}_{2\mu}=\bdtm+\tilde{b}_{2\mu}+a_{q}\lbb \lb \bdt \otimes \bdt \rb_{2\mu}+
\lb \bt \otimes \bt  \rb_{2\mu} \rbb,
\ee
with $a_{q}$ the anharmonic strength. Its reduced matrix elements on the states of the core are

\bea
\langle \pjgo || q_{0}^{c}Q^{c}_{2} || \pjgt \rangle &=&\frac{q_{eff}d}{\hat{J}_{2}\njgo\njgt}
\langle J_{1} 2; 00 | J_{2} 0 \rangle \times \\
\non &\times&\lbb \hat{J}^{2}_{1}\lb \njgo \rb^{2}+\hat{J}_{2}^{2} \lb \njgt \rb^{2} \rbb,
\eea
with $q_{eff}$ given by a linear formula in $d$

\be
q_{eff}=q_{0}^{c}\lb 1 - \sqrt{\frac{2}{7}}a_{q}d \rb.
\ee

The single particle quadrupole transition operator has the occupation number representation

\be
Q^{sp}_{2\mu}=\sum_{j_{1}j_{2}}\frac{1}{\hat{2}}\langle j_{1} || r^{2}Y_{2} || j_{2} \rangle 
\lb c^{\dagger}_{j_{1}} \tilde{c}_{j_{2}} \rb_{2\mu}.
\ee

Explicit expressions for the matrix elements of these operators over the states of the odd-mass nucleus
follow from the above results and the use of standard angular momentum algebra. 
For our particular case of fixed $j$, the final formulas are

\bea
\langle I_{1} || q_{0}^{c}Q^{c}_{2}|| I_{2}\rangle&=& \sum\limits_{J_{1}J_{2}}X_{I_{1}}^{J_{1}j}X_{I_{2}}^{J_{2}j}
\hat{I_{1}}\hat{I_{2}} \lb - \rb^{j-I_{1}} \times \\ 
\non &\times&\mathcal{W}\lb I_{1} J_{1} I_{2} J_{2};j2 \rb \langle \pjgo || q_{0}^{c} Q^{c}_{2}|| \pjgt \rangle, \\
\langle I_{1} || q_{0}^{sp}Q^{sp}_{2}|| I_{2}\rangle&=& \sum\limits_{J_{1}}X_{I_{1}}^{J_{1}j}X_{I_{2}}^{J_{1}j}\hat{I_{1}}\hat{I_{2}}
\lb - \rb^{j+I_{2}} \times \\
\non &\times& \mathcal{W}\lb I_{1} j I_{2} j;J_{1}2 \rb \langle j ||q_{0}^{sp}r^{2}Y_{2}|| j \rangle,
\eea
with $\mathcal{W}$ a Racah coefficient.

All reduced matrix elements are defined in the usual convention

\be
\langle l m| T_{\lambda \mu} | l' m' \rangle=\frac{1}{\hat{l}}\langle l'm' ; \lambda\mu | l m \rangle
\langle l||T_{\lambda}||l' \rangle.
\ee

\subsection{$\alpha$-emission in the coupled channels approach}

The decay phenomenon of interest connects the ground state of the parent nucleus of angular momentum $I_{P}$ to an excited
state of angular momentum $I$ of the daughter and an $\alpha$-particle of angular momentum $l$, in such a way
that the total angular momentum $I_{P}$ is conserved:

\be
P\lb I_{P} \rb \rightarrow D\lb I \rb + \alpha\lb l \rb.
\ee
An important remark is that both the initial state of the parent and the final state of the daughter
are built upon the same single particle orbital $j$.
This is known as a favored $\alpha$-transition, due to the fact that it usually has a large
branching ratio. The situation where the initial and final single particle orbitals are different
is known as an unfavored $\alpha$-transition. For the favored case, the transition from 
the ground state to the bandhead built atop the $j$ orbital in the daughter nucleus generally
has the highest decay width, and transitions on excited states of the band form the fine structure
of the spectrum.

The total wavefunction of the $\alpha$-daughter system can be assumed to be separable in radial
and angular parts and expanded in the angular momentum basis

\be\label{psiex}
\Psi \lb \bdt, {\bf R} \rb=\sum_{Il}\frac{f_{Il}\lb R \rb}{R}\mathcal{Z}_{Il}\lb \bdt, \omega \rb,
\ee
where the angular components are given by the coupling to good angular momentum 
between a wave function for the odd-mass daughter nucleus and a spherical
harmonic for the $\alpha$-particle

\be
\mathcal{Z}_{Il}\lb \bdt, \omega \rb=\lbb \Phi_{IM}\lb \bdt \rb \otimes Y_{lm} \lb \omega \rb \rbb_{I_{P}M_{P}}.
\ee
Here, ${\bf R}=\lb R, \omega \rb$ is the relative vector between the two fragments. Each pair of angular momentum values
defines a decay channel

\be
\lb I,l \rb=c.
\ee 

The function $\Psi$ must satisfy the stationary Schr\"{o}dinger equation

\be
H\Psi \lb \bdt,{\bf R} \rb=Q_{\alpha}\Psi \lb \bdt,{\bf R} \rb,
\ee
with $Q_{\alpha}$ the Q-value of the decay process. The Hamiltonian

\be
H=-\frac{\hbar}{2\mu}{\bf \nabla}^{2}_{{\bf R}} + H_{D} \lb \bdt \rb + V \lb \bdt, {\bf R} \rb
\ee
features a kinetic operator depending on the reduced mass $\mu$ of the system

\be
\mu=m_{N}\frac{4A_{D}}{4+A_{D}},
\ee
a term describing the motion of the daughter $H_{D}\lb \bdt \rb$ and an $\alpha$-daughter interaction
with monopole and quadrupole components

\be\label{vint}
V\lb \bdt,{\bf R} \rb=V_{0}\lb R \rb + V_{2} \lb \bdt,{\bf R} \rb.
\ee

A detailed study of this potential can be found in Ref. \cite{Del06}. There it is
shown that the monopole component has a pocket-like shape

\bea\label{vmon}
V_{0}\lb R\rb&=&v_{a}\bar{V}_{0} \lb R\rb, R > R_{m} \\
\non &=&a\lb R-R_{min} \rb^{2}-v_{0}, R<R_{m},
\eea
obtained through the matching of a harmonic oscillator to the
nuclear plus Coulomb potential $\bar{V}_{0}$ obtained by the method
of the double folding procedure of the M3Y particle-particle interaction
with Reid soft core parametrisation (Refs. \cite{Ber77,Sat79,Car92} and the book \cite{Del10}
for computational details).

The number $v_{a}$ acts as a quenching factor of the nuclear force. $v_{a}=1$ implies
an $\alpha$-particle existing with certainty, and a value $v_{a}<1$ is required in order
to simulate the formation of the $\alpha$-particle on the nuclear surface. Since branching ratios
tend to have a weak dependence on this parameter \cite{Del06}, it can be adjusted in order
to reproduce the total decay width $\Gamma$ \cite{Pel08}. Another possibility is to leave the interaction
potential unquenched and to consider the spectroscopic factor 

\be\label{specf}
S=\frac{\Gamma_{expt}}{\Gamma_{theor}},
\ee
as a measure of the particle formation probability, as in Ref. \cite{Del15a}.

The repulsive core on the second line of equation (\ref{vmon}) simulates the
Pauli principle, namely the fact that the $\alpha$-particle can exist only
on the nuclear surface. Its parameters can be adjusted so that the first
resonance in the potential corresponds to the experimental Q-value. 

The matching radius $R_{m}$ and the point $R_{min}$ at which the oscillator potential attains
the lowest value are found through the method of Ref. \cite{Del06}, which requires the
equality between the external attraction and internal repulsion together with their derivatives.
This makes the total interaction continuous and dependent on the repulsive strength $a$
and potential depth $v_{0}$. In our study, $a$ has a fixed value of $50\textrm{ MeV}$ for all nuclei and $v_{0}$
is fitted in each case through the experimental Q-value.

The second term of Eq. (\ref{vint}) is the QQ interaction

\bea
V_{2}\lb \bdt, {\bf R} \rb&=&-C_{0}\lb R-R_{min} \rb\frac{dV_{0}\lb R \rb}{dR}\times \\
\non &\times&\sqrt{5}\lbb Q^{c}_{2}\otimes Y_{2}\lb \omega \rb \rbb_{0},
\eea
with $C_{0}$ serving as an $\alpha$-nucleus coupling strength.

The angular functions entering the expansion of Eq. (\ref{psiex}) are orthonormal.
Using this, one obtains in a standard way the system of coupled differential equations
for radial components

\be\label{coupsys}
\frac{d^{2}f_{I_{1}l_{1}}\lb R \rb}{d\rho_{I_{1}}^{2}}=\sum_{I_{2}l_{2}}A_{I_{1}l_{1};I_{2}l_{2}}\lb R \rb f_{I_{2}l_{2}}\lb R \rb,
\ee
with the coupling matrix having the expression

\bea
\non A_{I_{1}l_{1};I_{2}l_{2}}\lb R \rb&=&\lbb \frac{l_{1}\lb l_{1}+1\rb}{\rho_{I_{1}}^{2}}+\frac{V_{0}\lb R \rb}{Q_{\alpha}-E_{I_{1}}}
-1 \rbb\delta_{I_{1}l_{1};I_{2}l_{2}}+\\
 &+&\frac{1}{Q_{\alpha}-E_{I_{1}}}\langle \mathcal{Z}_{I_{1}l_{1}}|V_{2}\lb \bdt,{\bf R} \rb| \mathcal{Z}_{I_{2}l_{2}}\rangle,
\eea
in terms of the reduced radius

\be\label{redrad}
\rho_{I}=\kappa_{I}R,\textrm{ }\kappa_{I}=\sqrt{\frac{2\mu \lb Q_{\alpha}-E_{I} \rb}{\hbar^{2}}}.
\ee
Notice that $\kappa_{I}$ has the same value for all the channels of fixed $I$, so the
supplementary $l$-index can be omitted both for the wave number and reduced radius.

The coupling term of the matrix is found by the same methods as in the previous sections to be

\bea
&&\non \langle \mathcal{Z}_{I_{1}l_{1}}|V_{2}\lb \bdt,{\bf R} \rb |\mathcal{Z}_{I_{2}l_{2}}\rangle=
\nn
&&\sum\limits_{J_{1}J_{2}}X_{I_{1}}^{J_{1}j}X_{I_{2}}^{J_{2}j}\langle \pjgo ||Q_{2}^{c}|| \pjgt \rangle
\langle l_{1} || Y_{2} || l_{2} \rangle
\hat{I}_{P}^{2}\hat{I}_{1}\hat{I}_{2}\hat{j}
\\&\times&
\lb - \rb ^{I_{2}-I_{P}+l_{2}}
\mathcal{W}\lb I_{1}l_{1}I_{2}l_{2};I_{P}2 \rb
\begin{Bmatrix} 
J_{1} & I_{1} & j \\ 
J_{2} & I_{2} & j \\ 
2 & 2 & 0  
\end{Bmatrix}
\eea
where the curly brackets denote a 9j-symbol. Since the reduced matrix element between the states of the
core is a linear function of the deformation \cite{Rad82}, one can express this linearity in terms of an effective
$\alpha$-nucleus coupling strength having a different anharmonic parameter $a_{\alpha}$

\be\label{ceff}
C=C_{0}\lb 1-\sqrt{\frac{2}{7}}a_{\alpha}d \rb.
\ee

\subsection{Resonant states}

The measured $\alpha$-decay widths are by many orders of magnitude smaller than the Q-value.
Thus, an $\alpha$-decaying state is almost a bound state, this being the main
reason way the stationary approach is a very good approximation of the emission process.
The state can be identified with a narrow resonant solution of 
the system of equations (\ref{coupsys}), containing only outgoing components.
In order to solve this system of equations we first define the internal and
external fundamental solutions which satisfy the boundary conditions

\bea
\label{orig}
&{\cal R}_{Il,L}(R)&
\stackrel{R\rightarrow 0}{\longrightarrow}
\delta_{Il,L}\varepsilon_{Il}~,
\nn
&{\cal H}_{Il,L}^{(+)}(R)&\equiv{\cal G}_{Il,L}(R) + 
i{\cal F}_{Il,L}(R)
\stackrel{R\rightarrow\infty}{\longrightarrow}
\nn &\delta_{Il,L}H_{l}^{(+)}(\kappa_{I} R)&
\equiv\delta_{Il,L}\left[G_{l}(\kappa_{I} R) + iF_{l}(\kappa_{I} R)\right]~,
\eea
where $\varepsilon_{Il}$ are arbitrary small numbers.
Here, the channel indexes label the component and $L$ the solution,
$G_{Il}(\kappa_{I} R)$ and $F_{Il}(\kappa_{I} R)$ are the standard irregular and regular
spherical Coulomb functions, depending on the
momentum $\kappa_{I}$ in the channel $c$, defined by Eq. (\ref{redrad}).

Each component of the solution is built as a superposition of $N$ independent
fundamental solutions.
We impose the matching conditions at some radius $R_1$ inside the barrier
and obtain
\bea
\label{match2}
f_{Il}(R_1)&=&\sum_L{\cal R}_{Il,L}(R_1)M_{Il,L}=\sum_L {\cal H}^{(+)}_{Il,L}(R_1) N_{Il,L}
\nn
\frac{df_{Il}(R_1)}{dR}&=&\sum_L\frac{d{\cal R}_{Il,L}(R_1)}{dR}M_{Il,L}=
\sum_L \frac{d{\cal H}^{(+)}_{Il,L}(R_1)}{dR} N_{Il,L}~,
\eea
where $N_{Il,L}$ are called scattering amplitudes.
One thus arrives at the following secular equation
\bea
\label{res}
&&\left|\begin{matrix}
{\cal R}(R_1) & {\cal H}^{(+)}(R_1) \cr
\frac{d{\cal R}(R_1)}{dR} & \frac{d{\cal H}^{(+)}(R_1)}{dR} \cr
\end{matrix}\right|
\approx
\left|\begin{matrix}
{\cal R}(R_1) & {\cal G}(R_1) \cr
\frac{d{\cal R}(R_1)}{dR} & \frac{d{\cal G}(R_1)}{dR} \cr
\end{matrix}\right|
=0~.
\eea
The first condition is fulfilled for the complex energies of the resonant states.
They practically coincide with the real scattering resonant states, 
due to the fact that the imaginary parts of energies are much smaller
than the corresponding real parts, which implies vanishing regular Coulomb
functions $F_{Il}$ inside the barrier.
The roots of the equation (\ref{res}) do not depend upon the matching radius $R_1$,
because both internal and external solutions satisfy the same Schr\"odinger equation.
The unknown coefficients $M_{Il,L}$ and $N_{Il,L}$ are obtained from
the normalization of the wave function in the internal region
\bea
\sum_{Il}\int_{R_0}^{R_2}\vert f_{Il}(R) \vert^2 dR =1~,
\eea
where $R_2$ is the external turning point.

From the continuity equation, the total decay width can be written as a sum of partial widths

\bea
\Gamma&=&\sum_{Il}\Gamma_{Il}=\sum_{Il}\hbar v_{I}\lim_{R\rightarrow \infty}|f_{Il}\lb R \rb|^{2}=\\
\non  &=&\sum_{Il}\hbar v_{I}|N_{Il}|^{2},
\eea
with $v_{I}$ the centre-of-mass velocity at infinity in the given channel

\be
v_{I}=\frac{\hbar\kappa_{I}}{\mu}.
\ee

\section{Numerical Application}

\begin{figure} 
\begin{center} 
\includegraphics[width=9cm]{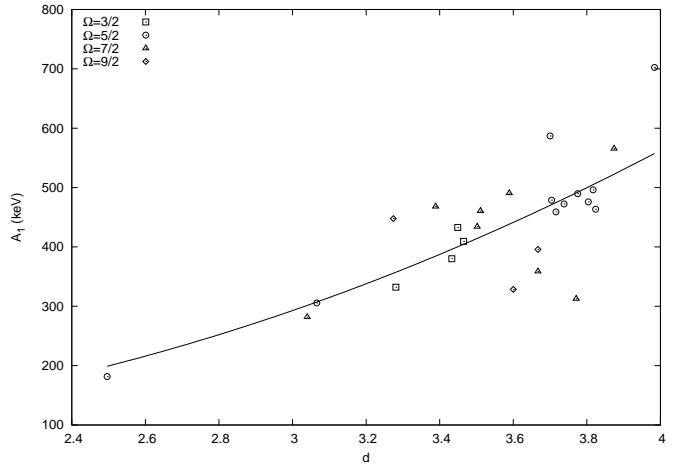}
\vspace{-8mm}
\caption{Hamiltonian strength parameter $A_{1}$ versus deformation $d$ for rotational bands built atop different values of
the odd nucleon angular momentum projection $\Omega$.}
\label{fig2}
\end{center} 
\end{figure}

All the experimental data with which we have tested the model has been provided by the ENSDF data set maintained by BNL \cite{Bnl}.
In this paper we have studied favored transitions in 26 odd-mass $\alpha$-emitters where the rotational band in which
the parent decays is built atop a single particle orbital of angular momentum projection $\Omega\ne \frac{1}{2}$.
Additionally, this band must be described in the formalism of an odd nucleon coupled to good angular momentum with a CSM core. The deformation
parameter $d$ was obtained by fitting available energy levels relative to the bandhead.
A number of about 4 levels is required for the determination of a reliable deformation. As can be seen from Fig. \ref{fig1}, 
there exists a deformation range where a large shift of the parameter's value has little impact on the energy levels. Because 
of this, when fewer energy levels are available, the fit becomes unreliable. In these circumstances we have determined the 
deformation parameter by studying the systematics of energy levels and deformations for the neighboring nuclei with good 
experimental data.
A quadratic trend is observed in the dependence of the Hamiltonian
stregth parameter $A_{1}$ on the deformation, as evidenced in Fig. \ref{fig2}, where we assign the nuclei with separate symbols
for each value of $\Omega$ . The fitting formula is

\bea
A_{1}\lb d \rb&=&55.583d^{2}-119.283d+150.409\\
\non \sigma&=&68.410,
\eea
agreeing qualitatively with the similar treatment made for the ground bands of even-even nuclei in Ref. \cite{Del15}.

\begin{figure}
\begin{center} 
\includegraphics[width=9cm]{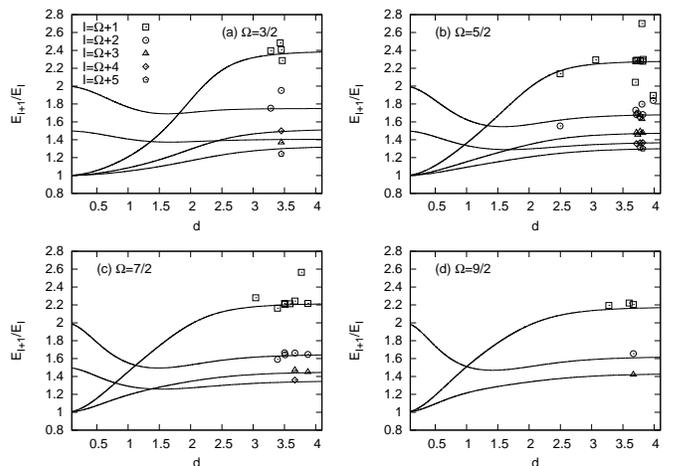}
\vspace{-8mm}
\caption{
Experimental energy level ratios $\frac{E_{I+1}}{E_{I}}$ as a function of the deformation parameter $d$ together with the
theoretical curves, separately for each value of the single particle angular momentum projection $\Omega$.}
\label{fig3}
\end{center} 
\end{figure}

The agreement between the ratio of experimental energy levels assigned to the deformation parameter $d$ and
the theoretical ratio $\frac{E_{I+1}}{E_{I}}$ is shown in Fig. \ref{fig3}, with separate panels 
for different values of $\Omega$.

On the topic of electromagnetic transitions, one notices a surprising lack of measured $B(E2)$ values
for odd-mass $\alpha$-emitters. Only one such value can be found in the database, for the transition
$\frac{9}{2}^{+}\rightarrow \frac{5}{2}^{+}$ in the ground band of $\textrm{Th}_{229}$. 
It is given by

\be 
B \lb E2;\frac{9}{2}^{+}\rightarrow \frac{5}{2}^{+} \rb=170\textit{ 30}\textrm{ W.u.}.
\ee
Using the systematics for the collective effective charge $q_{0}^{c}$ as function of $d$ established in Ref. \cite{Del15}
our model predicts a value

\be 
B\lb E2;\frac{9}{2}^{+}\rightarrow \frac{5}{2}^{+} \rb=117.8\textrm{ W.u.}. 
\ee
The difference up to the experimental value
can be obtained by tweaking the value of the single particle effective charge $q_{0}^{sp}$, which in this case must be
equal to $q_{0}^{sp}=7.004\textrm{ (W.u.)}^{\frac{1}{2}}$. Due to the lack of experimental data, a systematics of single particle
effective charges cannot currently be made, but we present predictions for $B\lb E2;\Omega+2 \rightarrow \Omega \rb$ 
values based on the systematics
of the collective effective charge from Ref. \cite{Del15}, together with results concerning energy levels in Table I.

To study $\alpha$-transitions, we make use of the so-called decay intensities

\be
\Upsilon_{Il}=\log_{10}\frac{\Gamma_{\Omega 0}}{\Gamma_{Il}},
\ee
and we will employ the notation $\Upsilon_{i},\textrm{ }i=1,2,3$ to refer to decay intensities
for the transitions to the first, second and third excited state respectively in any
rotational band of bandhead angular momentum projection $\Omega\ne \frac{1}{2}$. Notice that, in principle,
each intensity $\Upsilon_{i}$ is given by the sum

\be
\Upsilon_{i}=\sum_{l}\Upsilon_{Il},
\ee
where $I$ is fixed by the angular momentum of the daughter nucleus in that particular state and $l$ follows
from the triangle rule for the coupling to total angular momentum $I_{P}$. However, it is sufficient to consider
only one $l$-value for each state. This is due to the fact that the standard penetrability $P_{Il}$ through
the Coulomb barrier, defined by the factorization

\be\label{gamfac}
\Gamma_{Il}=2P_{Il}\lb R \rb \gamma^{2}_{Il}\lb R \rb,
\ee
decreases by one order of magnitude for each increasing value of $l$. Therefore, one would expect to be able
to make a reasonable prediction of the fine structure of the $\alpha$-emission spectrum using a basis of just four states, one state for the
bandhead and an additional state for each excited energy level. In the cases where experimental data concerning the energy
of the last state was not available, we used the CSM core + particle prediction provided by the fit of the lower energies.

\begin{figure} 
\begin{center} 
\includegraphics[width=9cm]{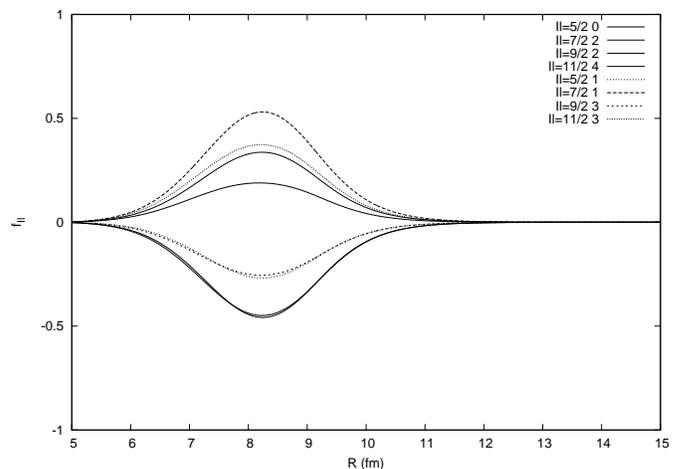}
\vspace{-8mm}
\caption{
Solutions to the system (\ref{coupsys}) for the favored decay process 
$\textrm{U}_{92}^{233}\rightarrow \textrm{Th}_{92}^{229} + \alpha_{2}^{4}$. Solid lines represent radial functions of even orbital angular 
momentum $l$ while dashed lines represent radial functions of odd $l$. The sets of functions of fixed parity are obtained simultaneously
for the same reaction energy and QQ coupling strength.}
\label{fig4}
\end{center} 
\end{figure}

It turns out however that the basis suggested above needs to be enlarged, due tot the fact that the parity of a resonance is fixed by whether
the $l$-values involved are even or odd. Since the interaction
(\ref{vint}) conserves parity, one must construct separate resonances of fixed even or odd parity. The even one follows the sequence of
minimal $l$-values in each channel as $l=0,2,2,4$, while the odd one follows the sequence $l=1,1,3,3$. Thus, each basis of four states
having a given parity constructs a separate resonant solution of the system (\ref{coupsys}). It is important
that both resonances are found at the same reaction energy $Q_{\alpha}$ and same QQ coupling stregth $C$. 
It is possible to achieve this for the potential of Eq. (\ref{vmon}) by adjusting the depth $v_{0}$ so that both resonances
generated at the same $C$ match in terms of the reaction energy. Using this, one can then tweak 
the effective coupling strength $C$ of Eq. (\ref{ceff}) to simultaneously generate different sets of even and odd resonances for each
$\alpha$-decay process of energy $Q_{\alpha}$, in an attempt to fit experimental data. 
One will thus obtain a total of eight radial functions in the solution, four in each resonance, as can be seen in Fig. \ref{fig4} for
the decay process 

\be
\textrm{U}_{92}^{233} \rightarrow \textrm{Th}_{92}^{229} + \alpha_{2}^{4}.
\ee

We have observed that for 23 decay processes out of the total of 26 studied, $C$ can be tweaked in order to match
the experimental value of $\Upsilon_{1}$ for a decay width with $l=0$ corresponding to the $\alpha$-transition to the bandhead and the
first decay width having $l=2$ obtained in the even resonance corresponding to the $\alpha$-transition to the first excited state. 
Simultaneously, the ratio between decay widths corresponding
to the same $l=0$ for the decay to the bandhead and the first value of $l=3$ for the decay to the second excited state obtained in the odd resonance
yielded a very good estimate of $\Upsilon_{2}$, while the ratio between decay widths corresponding to $l=0$ for the bandhead decay
and $l=4$ for the decay to the third excited state found in the even resonance
have given a reasonable value for $\Upsilon_{3}$. One of the exceptions is the decay to the daughter nucleus
$\textrm{Am}_{95}^{241}$, where the available data concerning $\Upsilon_{i},i=1,2$ suggests a doublet structure in the emission spectrum that
can be reproduced by employing the same $l=0$ width for the bandhead transition and the two decay widths with $l=2$ obtained in the even resonance.
The other exception concerns the two $\textrm{Ac}$
isotopes in our data set. In these cases, the decay width of angular momentum $l=0$ and the first $l=2$ width obtained in the even resonance can be
used to reproduce the value of $\Upsilon_{2}$, situation in which the $l=0$ width and the second width of angular momentum $l=1$
in the odd resonance (which corresponds to the transition to the first excited state) will reproduce $\Upsilon_{1}$ reasonably. 

\begin{figure} 
\begin{center} 
\includegraphics[width=9cm]{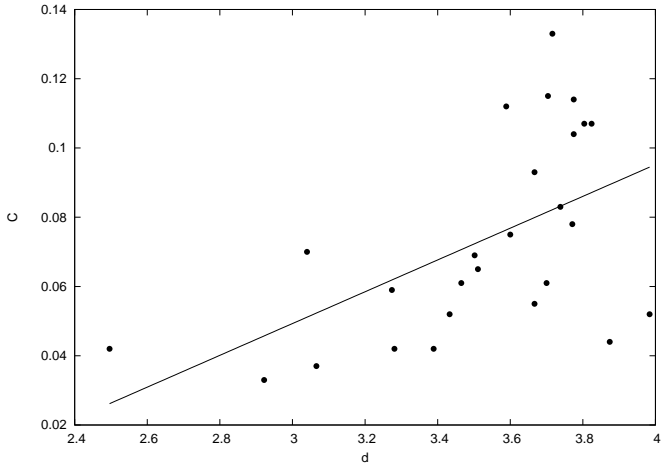}
\vspace{-8mm}
\caption{
Effective $\alpha$-nucleus coupling strength $C$ versus deformation parameter $d$.}
\label{fig5}
\end{center} 
\end{figure}

When plotted against the deformation parameter,
the values of $C$ obtained from the above fit follow the prediction of Eq. (\ref{ceff}) by exhibiting a linear trend with respect to $d$,
as seen in Fig. \ref{fig5}. The parameters of the linear fit are
\be
C_{0}=-0.088,\textrm{ }a_{\alpha}=0.971,\textrm{ }\sigma=0.023.
\ee

\begin{figure}
\begin{center} 
\includegraphics[width=9cm]{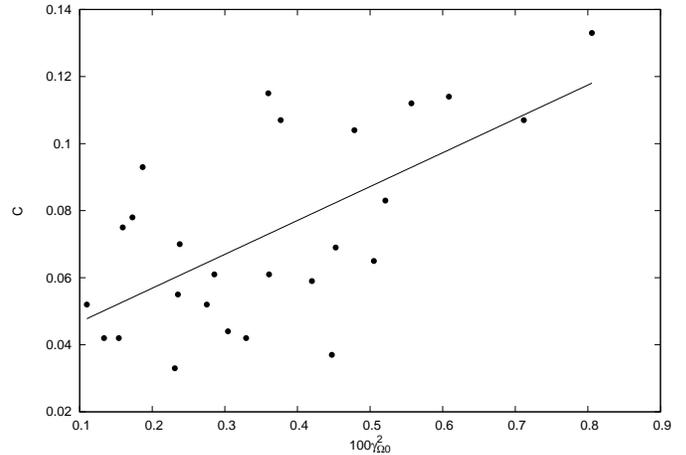}
\vspace{-8mm}
\caption{
Effective $\alpha$-nucleus coupling strength $C$ versus the reduced width $\gamma^{2}_{\Omega0}$ for $\alpha$-transitions
to the bandhead.}
\label{fig6}
\end{center} 
\end{figure}

This coupling strength can be interpreted as a measure of $\alpha$-clustering. To see this, we
use the reduced width $\gamma^{2}_{\Omega 0}$ introduced in Eq. (\ref{gamfac}).
It turns out that $C$ shows a linear correlation with $\gamma^{2}_{\Omega0}$ with a positive slope,
as can be seen in Fig. \ref{fig6}. The parameters are given by

\bea\label{credl}
C&=&10.096\gamma^{2}_{\Omega}+0.037 \\
\non \sigma&=&0.021.
\eea

\begin{figure}
\begin{center} 
\includegraphics[width=9cm]{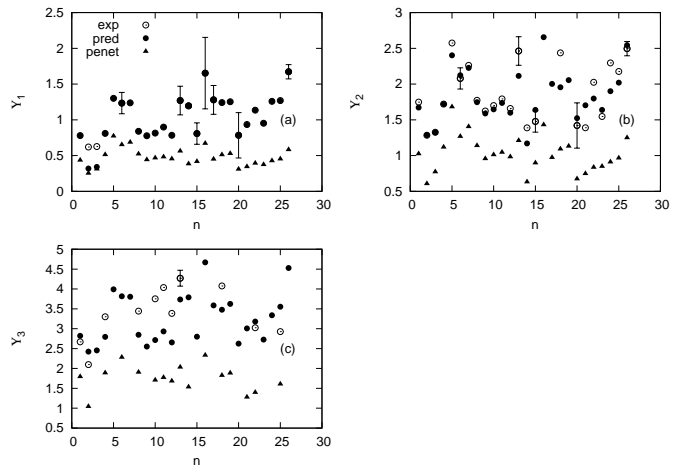}
\vspace{-8mm}
\caption{
Intensities of the favored $\alpha$-transitions $\Upsilon_{i}$ 
to the first three excited states in rotational bands as function of the index number $n$
in the first column of Tables I and II.
Open circles denote the experimental data,
filled circles are the values predicted by the coupled channels method with a particle + CSM core
structure model and dark triangles show the barrier penetration estimates.}
\label{fig7}
\end{center} 
\end{figure}

In Fig. \ref{fig7} we present in separate panels the values of the intensities 
$\Upsilon_{i},\textrm{ }i=1,2,3$ obtained by the method presented above, versus the index number $n$ found in the first column of Tables I and II.
With open circles we show experimental data and 
with filled circles we give the values predicted by the coupled channels method
with a particle + CSM core structure model. Dark triangles present the crudest barrier penetration calculation,
where the intensities follow from the ratios of penetrabilities computed at the same values of $l$ as in the
coupled channels approach

\be
\Upsilon_{i}=log_{10}\frac{P_{\Omega 0}}{P_{I l}}.
\ee
All emission data is presented in Table II.

\begin{figure}
\begin{center} 
\includegraphics[width=9cm]{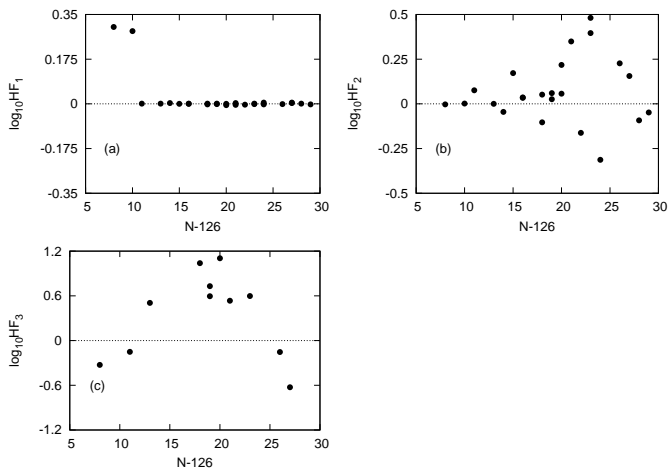}
\vspace{-8mm}
\caption{
Logarithm of the hindrance factor $HF_{i}$ versus neutron number $N-126$, separately for each excited state $i=1,2,3$.}
\label{fig8}
\end{center} 
\end{figure}

As we mentioned, the spectroscopic factor defined by Eq. (\ref{specf})
accounts for clustering effects. One can define partial spectroscopic
factors for each channel and the logarithm of the hindrance factor as
\be
\log_{10}HF_{Il}=\log\frac{S_{\Omega 0}}{S_{Il}}=
\Upsilon_{Il}^{exp}-\Upsilon_{Il}^{theor}.
\ee
This quantity shows the importance of the extra-clustering in the decay process to excited states that is
not considered within our model. 
In Fig. \ref{fig8}
we have plotted these logarithms versus the neutron number. It is clearly shown that coupling an $\alpha$-particle
to the daughter nucleus with the required strength needed to reproduce one intensity (usually $\Upsilon_{1}$, with
the exception of Ac isotopes where $\Upsilon_{2}$ is reproduced) allows one
to predict the values of the other intensities within a factor usually less than 3.

\begin{figure}
\begin{center} 
\includegraphics[width=9cm]{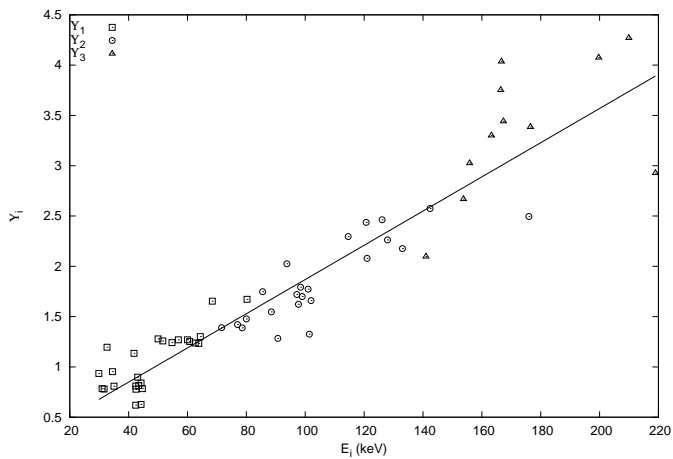}
\vspace{-8mm}
\caption{
$\Upsilon_{i}$ values versus excitation energy $E_{i}$ relative to the bandhead in each case.}
\label{fig9}
\end{center} 
\end{figure}

We note that the universal decay law treated in Refs. \cite{Del15a} and \cite{Del09} is once again manifested
in the dependence of the decay intensities on excitation energies. In Fig. \ref{fig9} we have represented
all of the $\Upsilon_{i}$ values as function of the corresponding excitation energy $E_{i}$ relative to the bandhead
for each collective structure analyzed
in this paper. We observe a universal linear correlation with parameters

\be
\Upsilon_{i}=0.017E_{i}+0.169,~\sigma=0.316.
\ee

\begin{figure}
\begin{center} 
\includegraphics[width=9cm]{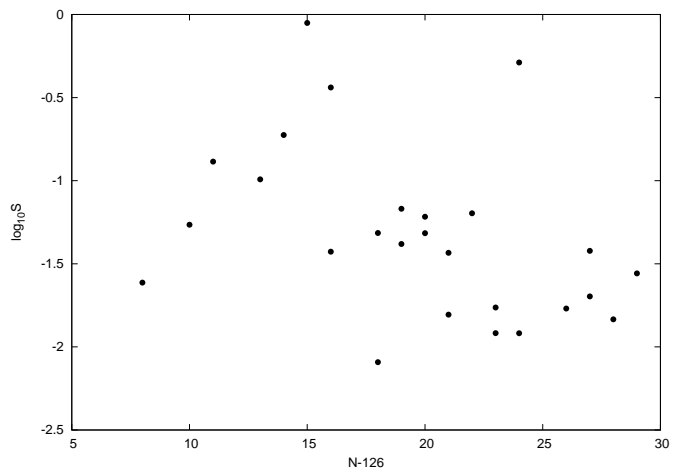}
\vspace{-8mm}
\caption{
Logarithm of the spectroscopic factor $S$ versus neutron number $N-126$.}
\label{fig10}
\end{center} 
\end{figure}

As a final remark, the logarithm of the spectroscopic factor of Eq. (\ref{specf}) can be represented as a function of neutron number,
like in Fig. \ref{fig10}. This quantity shows a decreasing trend with the neutron number, meaning that the unquenched potential predicts
shorter half-lives for heavier nuclei than what is observed experimentally.

\section{Conclusions}

We analyzed the available experimental data for favored $\alpha$-transitions to rotational bands built upon a single particle
angular momentum projection $\Omega\ne \frac{1}{2}$. The nuclear structure was modeled as an odd-nucleon coupled to a coherent state
even-even core, the energy levels of each band being fitted through the use of a deformation parameter $d$
 and Hamiltonian strength parameter $A_{1}$ that is related to the deformation through a quadratic dependence. 
$B\lb E 2 \rb$ values can be predicted using the systematics of the collective effective charge
as function of deformation established in Ref. \cite{Del15}. In the absence of experimental data that allows the study of the
single particle effective charge contribution, it is expected that these predicted values are smaller than what will be observed in reality. 

The fine structure of the $\alpha$-emission spectrum was studied using the coupled channels method, through a QQ interaction tweaked
by a coupling strength that behaves linearly with respect to the deformation parameter and reduced width for the
g.s.$\rightarrow \Omega $ transition.
The predicted values of the intensities are in reasonable agreement with experimental data, usually within a factor
less than 3. With additional developments in the structure part, it is expected that the model will be useful for the
study of the case $\Omega=\frac{1}{2}$ as well.  

\begin{acknowledgments}

This work was supported by the grants of the Romanian Ministry of Education and Research,
CNCS – UEFISCDI, PN-II-ID-PCE-2011-3-0092, PN-09370102 and
by the strategic grant POSDRU/159/1.5/S/137750.

\end{acknowledgments}

\clearpage 

\begin{longtable}{cccccccc}
\caption{Bandhead spin and parity, deformation and 
Hamiltonian strength parameters,
experimental and
predicted excited energy levels, $B(E2)$ values for the transition
$\Omega+2\rightarrow \Omega$ for
rotational bands in daughter nuclei
where favored $\alpha$-transitions occur.
\label{tab1}}\\
\hline
\hline
 n & $D\lb I \rb $ & $\Omega^{\pi}$ & $d$ & $A_{1}$ & $E_{exp}$ & $E_{fit}$ & $B\lb E2\rb_{\Omega+2 \rightarrow \Omega} $ \cr
   &              &                &     & keV     &  keV      & keV       & W.u. \cr
\hline
   1 &   Ra$_{88}^{225}$ &    5$^{+}$ &      3.804 &    475.876 &    236.25$\textit{  3}$ &   - &    124.244 \cr
     &            &           7$^{+}$ &            &            &    267.92$\textit{  5}$ &    285.044 & \cr
     &            &           9$^{+}$ &            &            &    321.76$\textit{  8}$ &    336.732 & \cr
     &            &          11$^{+}$ &            &            &    390.0$\textit{  4}$ &    399.098 & \cr
     &            &          13$^{+}$ &            &            &    487$\textit{  3}$ &    471.662 & \cr
     &            &          15$^{+}$ &            &            &                        - &    554.036 & \cr
     &            &          17$^{+}$ &            &            &                        - &    645.556 & \cr
   2 &   Ac$_{89}^{223}$ &    5$^{-}$ &      2.496 &    181.721 &     0  &             - &     49.855 \cr
     &            &           7$^{-}$ &            &            &            42.4$\textit{  1}$ &     43.582 & \cr
     &            &           9$^{-}$ &            &            &     90.7$\textit{  1}$ &     89.180 & \cr
     &            &          11$^{-}$ &            &            &    141$\textit{  5}$ &    141.602 & \cr
     &            &          13$^{-}$ &            &            &                        - &    197.948 & \cr
     &            &          15$^{-}$ &            &            &                        - &    260.189 & \cr
     &            &          17$^{-}$ &            &            &                        - &    323.633 & \cr
   3 &   Ac$_{89}^{225}$ &    5$^{+}$ &      3.066 &    305.552 &    155.65$\textit{  7}$ &          - &     76.802 \cr
     &            &           7$^{+}$ &            &            &           199.85$\textit{  9}$ &    203.667 & \cr
     &            &           9$^{+}$ &            &            &    257.04$\textit{ 16}$ &    255.341 & \cr
     &            &          11$^{+}$ &            &            &                        - &    316.518 & \cr
     &            &          13$^{+}$ &            &            &                        - &    385.988 & \cr
     &            &          15$^{+}$ &            &            &                        - &    463.502 & \cr
     &            &          17$^{+}$ &            &            &                        - &    547.239 & \cr
   4 &   Th$_{90}^{229}$ &    5$^{+}$ &      3.716 &    458.930 &     0 &                 - &    117.801 \cr
     &            &           7$^{+}$ &            &            &            42.4349$\textit{  2}$ &     49.150 & \cr
     &            &           9$^{+}$ &            &            &     97.13595$\textit{ 24}$ &    101.467 & \cr
     &            &          11$^{+}$ &            &            &    163.2542$\textit{  7}$ &    164.504 & \cr
     &            &          13$^{+}$ &            &            &    241.546$\textit{ 19}$ &    237.728 & \cr
     &            &          15$^{+}$ &            &            &    327.8$\textit{  3}$ &    320.725 & \cr
     &            &          17$^{+}$ &            &            &                        - &    412.752 & \cr
   5 &   Th$_{90}^{231}$ &    7$^{-}$ &      3.589 &    490.657 &    387.836$\textit{ 1}$ &           - &     79.654 \cr
     &            &           9$^{-}$ &            &            &           452.176$\textit{ 15}$ &    457.046 & \cr
     &            &          11$^{-}$ &            &            &    530.24$\textit{  5}$ &    528.029 & \cr
     &            &          13$^{-}$ &            &            &                        - &    610.351 & \cr
     &            &          15$^{-}$ &            &            &                        - &    703.366 & \cr
     &            &          17$^{-}$ &            &            &                        - &    806.423 & \cr
     &            &          19$^{-}$ &            &            &                        - &    918.846 & \cr
   6 &   Pa$_{91}^{231}$ &    5$^{+}$ &      3.984 &    702.273 &    183.4962$\textit{ 17}$ &         - &    138.117 \cr
     &            &           7$^{+}$ &            &            &           247.320$\textit{  5}$ &    246.436 & \cr
     &            &           9$^{+}$ &            &            &    304.5$\textit{  4}$ &    315.768 & \cr
     &            &          11$^{+}$ &            &            &    406.1$\textit{  3}$ &    399.615 & \cr
     &            &          13$^{+}$ &            &            &                        - &    497.451 & \cr
     &            &          15$^{+}$ &            &            &                        - &    608.810 & \cr
     &            &          17$^{+}$ &            &            &                        - &    732.964 & \cr
   7 &   Pa$_{91}^{233}$ &    5$^{+}$ &      3.700 &    587.036 &    237.89$\textit{ 13}$ &          - &    116.653 \cr
     &            &           7$^{+}$ &            &            &           300.50$\textit{  3}$ &    298.987 & \cr
     &            &           9$^{+}$ &            &            &    365.84$\textit{  8}$ &    366.507 & \cr
     &            &          11$^{+}$ &            &            &                        - &    447.840 & \cr
     &            &          13$^{+}$ &            &            &    589$\textit{  4}$ &    542.285 & \cr
     &            &          15$^{+}$ &            &            &                        - &    649.306 & \cr
     &            &          17$^{+}$ &            &            &                        - &    767.923 & \cr
     & & & & & & & \cr
     & & & & & & & \cr
     & & & & & & & \cr
     & & & & & & & \cr
     & & & & & & & \cr
     & & & & & & & \cr
     & & & & & & & \cr
     & & & & & & & \cr
     & & & & & & & \cr
     & & & & & & & \cr
     & & & & & & & \vspace{0.71mm} \cr
\hline
\hline
 n & $D\lb I \rb $ & $\Omega^{\pi}$ & $d$ & $A_{1}$ & $E_{exp}$ & $E_{fit}$ & $B\lb E2\rb_{\Omega+2 \rightarrow \Omega} $ \cr
   &              &                &     & keV     &  keV      & keV       & W.u. \cr
\hline
   8 &   U $_{92}^{237}$ &    5$^{+}$ &      3.775 &    489.617 &    159.962$\textit{ 14}$ &         - &    122.096 \cr
     &            &           7$^{+}$ &            &            &           204.17$\textit{  7}$ &    210.550 & \cr
     &            &           9$^{+}$ &            &            &    260.93$\textit{ 12}$ &    264.578 & \cr
     &            &          11$^{+}$ &            &            &    327.3$\textit{ 10}$ &    329.739 & \cr
     &            &          13$^{+}$ &            &            &    409.8$\textit{ 10}$ &    405.515 & \cr
     &            &          15$^{+}$ &            &            &    501.4$\textit{ 12}$ &    491.493 & \cr
     &            &          17$^{+}$ &            &            &    607.7$\textit{ 12}$ &    586.958 & \cr
   9 &   Np$_{93}^{235}$ &    5$^{-}$ &      3.824 &    463.486 &     49.10$\textit{  5}$ &         - &    125.739 \cr
     &            &           7$^{-}$ &            &            &            91.6$\textit{  3}$ &     95.349 & \cr
     &            &           9$^{-}$ &            &            &    146.8$\textit{  7}$ &    145.150 & \cr
     &            &          11$^{-}$ &            &            &                        - &    205.255 & \cr
     &            &          13$^{-}$ &            &            &                        - &    275.213 & \cr
     &            &          15$^{-}$ &            &            &                        - &    354.654 & \cr
     &            &          17$^{-}$ &            &            &                        - &    442.954 & \cr
  10 &   Np$_{93}^{237}$ &    5$^{-}$ &      3.817 &    496.159 &     59.54092$\textit{ 22}$ &        - &    125.214 \cr
     &            &           7$^{-}$ &            &            &           102.959$\textit{  3}$ &    108.698 & \cr
     &            &           9$^{-}$ &            &            &    158.497$\textit{  3}$ &    162.212 & \cr
     &            &          11$^{-}$ &            &            &    225.957$\textit{ 16}$ &    226.791 & \cr
     &            &          13$^{-}$ &            &            &    305.050$\textit{  3}$ &    301.948 & \cr
     &            &          15$^{-}$ &            &            &    395.53$\textit{  4}$ &    387.283 & \cr
     &            &          17$^{-}$ &            &            &    497.01$\textit{  5}$ &    482.120 & \cr
  11 &   Np$_{93}^{239}$ &    5$^{-}$ &      3.738 &    472.328 &     74.6640$\textit{ 10}$ &         - &    119.391 \cr
     &            &           7$^{-}$ &            &            &           117.715$\textit{ 40}$ &    123.468 & \cr
     &            &           9$^{-}$ &            &            &    173.086$\textit{ 18}$ &    176.660 & \cr
     &            &          11$^{-}$ &            &            &    241.312$\textit{ 24}$ &    240.775 & \cr
     &            &          13$^{-}$ &            &            &    317.4$\textit{ 15}$ &    315.282 & \cr
     &            &          15$^{-}$ &            &            &                        - &    399.767 & \cr
     &            &          17$^{-}$ &            &            &                        - &    493.493 & \cr
  12 &   Pu$_{94}^{239}$ &    5$^{-}$ &      3.704 &    478.740 &    285.460$\textit{  2}$ &          - &    116.939 \cr
     &            &           7$^{-}$ &            &            &           330.124$\textit{  4}$ &    336.658 & \cr
     &            &           9$^{-}$ &            &            &    387.42$\textit{  2}$ &    391.599 & \cr
     &            &          11$^{-}$ &            &            &    462$\textit{  3}$ &    457.784 & \cr
     &            &          13$^{-}$ &            &            &                        - &    534.646 & \cr
     &            &          15$^{-}$ &            &            &                        - &    621.748 & \cr
     &            &          17$^{-}$ &            &            &                        - &    718.299 & \cr
  13 &   Pu$_{94}^{241}$ &    7$^{+}$ &      3.502 &    434.114 &    175.0523$\textit{ 14}$ &         - &     75.420 \cr
     &            &           9$^{+}$ &            &            &           231.935$\textit{  9}$ &    238.742 & \cr
     &            &          11$^{+}$ &            &            &    301.172$\textit{ 16}$ &    304.665 & \cr
     &            &          13$^{+}$ &            &            &    385$\textit{  3}$ &    380.962 & \cr
     &            &          15$^{+}$ &            &            &                        - &    466.987 & \cr
     &            &          17$^{+}$ &            &            &                        - &    562.094 & \cr
     &            &          19$^{+}$ &            &            &                        - &    665.617 & \cr
  14 &   Am$_{95}^{241}$ &    3$^{-}$ &      3.449 &    432.807 &    471.812$\textit{  9}$ &          - &    141.463 \cr
     &            &           5$^{-}$ &            &            &           504.451$\textit{  9}$ &    510.215 & \cr
     &            &           7$^{-}$ &            &            &    550.4$\textit{  4}$ &    556.342 & \cr
     &            &           9$^{-}$ &            &            &    625.2$\textit{  5}$ &    616.164 & \cr
     &            &          11$^{-}$ &            &            &    682.1$\textit{  6}$ &    684.941 & \cr
     &            &          13$^{-}$ &            &            &    787.2$\textit{  6}$ &    768.880 & \cr
     &            &          15$^{-}$ &            &            &    863.8$\textit{  7}$ &    856.494 & \cr
     & & & & & & & \cr
     & & & & & & & \cr
     & & & & & & & \cr
     & & & & & & & \cr
     & & & & & & & \cr
     & & & & & & & \cr
     & & & & & & & \cr
\hline
\hline
 n & $D\lb I \rb $ & $\Omega^{\pi}$ & $d$ & $A_{1}$ & $E_{exp}$ & $E_{fit}$ & $B\lb E2\rb_{\Omega+2 \rightarrow \Omega} $ \cr
   &              &                &     & keV     &  keV      & keV       & W.u. \cr
\hline
  15 &   Am$_{95}^{243}$ &    3$^{-}$ &      3.465 &    409.433 &    265$\textit{  10}$ &         - &    142.950 \cr
     &            &           5$^{-}$ &            &            &           300$\textit{  2}$ &    301.257 & \cr
     &            &           7$^{-}$ &            &            &    345$\textit{  1}$ &    344.467 & \cr
     &            &           9$^{-}$ &            &            &                        - &    400.496 & \cr
     &            &          11$^{-}$ &            &            &                        - &    464.992 & \cr
     &            &          13$^{-}$ &            &            &                        - &    543.654 & \cr
     &            &          15$^{-}$ &            &            &                        - &    625.919 & \cr
  16 &   Am$_{95}^{245}$ &    7$^{+}$ &      3.389 &    467.904 &    327.428$\textit{  8}$ &          - &     70.148 \cr
     &            &           9$^{+}$ &            &            &           395.870$\textit{  2}$ &    399.236 & \cr
     &            &          11$^{+}$ &            &            &    475.52$\textit{  3}$ &    475.021 & \cr
     &            &          13$^{+}$ &            &            &    563.1$\textit{  3}$ &    562.466 & \cr
     &            &          15$^{+}$ &            &            &                        - &    660.747 & \cr
     &            &          17$^{+}$ &            &            &                        - &    769.069 & \cr
     &            &          19$^{+}$ &            &            &                        - &    886.603 & \cr
  17 &   Cm$_{96}^{243}$ &    7$^{+}$ &      3.040 &    281.795 &    114$\textit{  20}$ &         - &     55.439 \cr
     &            &           9$^{+}$ &            &            &           164$\textit{  3}$ &    169.285 & \cr
     &            &          11$^{+}$ &            &            &    228$\textit{  3}$ &    225.576 & \cr
     &            &          13$^{+}$ &            &            &                        - &    289.679 & \cr
     &            &          15$^{+}$ &            &            &                        - &    360.773 & \cr
     &            &          17$^{+}$ &            &            &                        - &    438.184 & \cr
     &            &          19$^{+}$ &            &            &                        - &    521.137 & \cr
  18 &   Cm$_{96}^{245}$ &    9$^{-}$ &      3.667 &    395.789 &    388.181$\textit{ 13}$ &          - &     63.542 \cr
     &            &          11$^{-}$ &            &            &           442.918$\textit{ 21}$ &    453.613 & \cr
     &            &          13$^{-}$ &            &            &    508.87$\textit{  3}$ &    516.475 & \cr
     &            &          15$^{-}$ &            &            &    587.9$\textit{ 10}$ &    587.652 & \cr
     &            &          17$^{-}$ &            &            &    672$\textit{  3}$ &    666.679 & \cr
     &            &          19$^{-}$ &            &            &                        - &    753.084 & \cr
     &            &          21$^{-}$ &            &            &                        - &    846.399 & \cr
  19 &   Cm$_{96}^{249}$ &    7$^{+}$ &      3.511 &    460.624 &     48.76$\textit{  4}$ &          - &     75.851 \cr
     &            &           9$^{+}$ &            &            &           109.49$\textit{  9}$ &    115.521 & \cr
     &            &          11$^{+}$ &            &            &    182.77$\textit{ 16}$ &    185.116 & \cr
     &            &          13$^{+}$ &            &            &    268.8$\textit{  3}$ &    265.682 & \cr
     &            &          15$^{+}$ &            &            &                        - &    356.540 & \cr
     &            &          17$^{+}$ &            &            &                        - &    457.015 & \cr
     &            &          19$^{+}$ &            &            &                        - &    566.406 & \cr
  20 &   Bk$_{97}^{241}$ &    3$^{-}$ &      3.433 &    380.147 &     51$\textit{  4}$ &          - &    139.986 \cr
     &            &           5$^{-}$ &            &            &            82$\textit{  6}$ &     85.570 & \cr
     &            &           7$^{-}$ &            &            &    128$\textit{  7}$ &    126.486 & \cr
     &            &           9$^{-}$ &            &            &                        - &    179.561 & \cr
     &            &          11$^{-}$ &            &            &                        - &    240.500 & \cr
     &            &          13$^{-}$ &            &            &                        - &    314.926 & \cr
     &            &          15$^{-}$ &            &            &                        - &    392.457 & \cr
  21 &   Bk$_{97}^{247}$ &    3$^{-}$ &      3.281 &    332.176 &     0 &               - &    126.436 \cr
     &            &           5$^{-}$ &            &            &            29.88$\textit{ 11}$ &     33.443 & \cr
     &            &           7$^{-}$ &            &            &     71.60$\textit{ 13}$ &     72.792 & \cr
     &            &           9$^{-}$ &            &            &    125.5$\textit{  4}$ &    123.949 & \cr
     &            &          11$^{-}$ &            &            &                        - &    181.849 & \cr
     &            &          13$^{-}$ &            &            &                        - &    253.117 & \cr
     &            &          15$^{-}$ &            &            &                        - &    325.800 & \cr
  22 &   Bk$_{97}^{249}$ &    7$^{+}$ &      3.667 &    358.729 &     0 &                - &     83.581 \cr
     &            &           9$^{+}$ &            &            &            41.805$\textit{  8}$ &     50.471 & \cr
     &            &          11$^{+}$ &            &            &     93.759$\textit{  8}$ &    100.203 & \cr
     &            &          13$^{+}$ &            &            &    155.854$\textit{ 10}$ &    157.973 & \cr
     &            &          15$^{+}$ &            &            &    229.242$\textit{ 12}$ &    223.360 & \cr
     &            &          17$^{+}$ &            &            &    311.857$\textit{ 23}$ &    295.932 & \cr
     &            &          19$^{+}$ &            &            &                        - &    375.243 & \cr
     & & & & & & & \cr
     & & & & & & & \cr
     & & & & & & & \cr
     & & & & & & & \vspace{0.01mm} \cr
\hline
\hline
 n & $D\lb I \rb $ & $\Omega^{\pi}$ & $d$ & $A_{1}$ & $E_{exp}$ & $E_{fit}$ & $B\lb E2\rb_{\Omega+2 \rightarrow \Omega} $ \cr
   &              &                &     & keV     &  keV      & keV       & W.u. \cr
\hline
  23 &   Bk$_{97}^{251}$ &    7$^{+}$ &      3.771 &    312.780 &     $\sim$35.5 &                - &     89.011 \cr
     &            &           9$^{+}$ &            &            &            70$\textit{  3}$ &     78.934 & \cr
     &            &          11$^{+}$ &            &            &    $\sim$124 &    119.950 & \cr
     &            &          13$^{+}$ &            &            &                        - &    167.687 & \cr
     &            &          15$^{+}$ &            &            &                        - &    221.829 & \cr
     &            &          17$^{+}$ &            &            &                        - &    282.046 & \cr
     &            &          19$^{+}$ &            &            &                        - &    347.998 & \cr
  24 &   Cf$_{98}^{247}$ &    9$^{-}$ &      3.600 &    328.578 &    480.40$\textit{ 9}$ &           - &     61.000 \cr
     &            &          11$^{-}$ &            &            &           531.99$\textit{ 21}$ &    538.111 & \cr
     &            &          13$^{-}$ &            &            &    595$\textit{  4}$ &    592.167 & \cr
     &            &          15$^{-}$ &            &            &                        - &    653.285 & \cr
     &            &          17$^{-}$ &            &            &                        - &    721.042 & \cr
     &            &          19$^{-}$ &            &            &                        - &    795.013 & \cr
     &            &          21$^{-}$ &            &            &                        - &    874.780 & \cr
  25 &   Cf$_{98}^{251}$ &    7$^{+}$ &      3.874 &    565.534 &    106.309$\textit{ 18}$ &          - &     94.611 \cr
     &            &           9$^{+}$ &            &            &           166.303$\textit{ 23}$ &    174.178 & \cr
     &            &          11$^{+}$ &            &            &    239.33$\textit{  3}$ &    244.459 & \cr
     &            &          13$^{+}$ &            &            &    325.29$\textit{  3}$ &    326.389 & \cr
     &            &          15$^{+}$ &            &            &    423.92$\textit{  4}$ &    419.477 & \cr
     &            &          17$^{+}$ &            &            &                        - &    523.204 & \cr
     &            &          19$^{+}$ &            &            &                        - &    637.027 & \cr
  26 &   Cf$_{98}^{253}$ &    9$^{+}$ &      3.274 &    447.757 &    241.01$\textit{  8}$ &          - &     49.612 \cr
     &            &          11$^{+}$ &            &            &           321.21$\textit{ 22}$ &    326.500 & \cr
     &            &          13$^{+}$ &            &            &    417$\textit{  5}$ &    414.539 & \cr
     &            &          15$^{+}$ &            &            &                        - &    513.164 & \cr
     &            &          17$^{+}$ &            &            &                        - &    621.496 & \cr
     &            &          19$^{+}$ &            &            &                        - &    738.703 & \cr
     &            &          21$^{+}$ &            &            &                        - &    863.988 & \cr
\hline
\end{longtable}

\clearpage
\begin{center}
\begin{table}\label{tab2}
\caption{Sequence of $l$-values that reproduces the fine structure of the emission spectrum,
QQ coupling strength between $\alpha$-particle and odd-mass daughter nucleus, Q-value for the g.s.$\rightarrow\textrm{ }\Omega 0$ transition,
experimental and predicted logarithm of the total half-life in the $\alpha$-channel, 
experimental and predicted values for the favored decay
intensities to the first three excited states of each rotational band.}
\begin{tabular}{cccccccccccccccc}
\hline
\hline
  n & $D\lb I \rb$ & $l_{g.s. \rightarrow \Omega}$ & $l_{g.s. \rightarrow \Omega+1}$ & $l_{g.s. \rightarrow \Omega+2}$ & $l_{g.s. \rightarrow \Omega+3}$ & C & $Q_{g.s.\rightarrow \Omega}$ & $\log_{10}T_{\alpha}^{exp}$ & $\log_{10}T_{\alpha}^{pred}$ &   $\Upsilon_{1}^{exp}$ & $\Upsilon_{1}^{pred}$ & $\Upsilon_{2}^{exp}$ & $\Upsilon_{2}^{pred}$ & $\Upsilon_{3}^{exp}$ & $\Upsilon_{3}^{pred}$ \cr
    &              &   &   &   &   &   &   & (MeV)                    & (s)                         & (s)                         &  & & & & \cr
  1 &   Ra$_{88}^{225}$ & 0 & 2 & 3 & 4 &    0.107 &    4.931 & 11.362  &  10.477 &       0.780 &    0.779 &    1.748 &    1.672 &    2.669 &    2.820 \cr 
  2 &   Ac$_{89}^{223}$ & 0 & 1 & 2 & 3 &    0.073 &    6.580 &  3.432  &   1.819 &      0.620 &    0.319 &    1.284 &    1.287 &    2.097 &    2.424 \cr 
  3 &   Ac$_{89}^{225}$ & 0 & 1 & 2 & 3 &    0.085 &    5.679 &  7.433  &   6.168 &      0.627 &    0.342 &    1.326 &    1.324 &        - &    2.455 \cr 
  4 &   Th$_{90}^{229}$ & 0 & 2 & 3 & 4 &    0.133 &    4.909 &  12.669 &  11.677 &       0.810 &    0.809 &    1.720 &    1.720 &    3.301 &    2.795 \cr 
  5 &   Th$_{90}^{231}$ & 0 & 2 & 3 & 4 &    0.112 &    4.290 &  16.342 &  16.291 &       1.301 &    1.301 &    2.574 &    2.402 &        - &    3.990 \cr 
  6 &   Pa$_{91}^{231}$ & 0 & 2 & 3 & 4 &    0.052 &    5.011 &  12.117 &  11.392 &       1.234 &    1.231 &    2.079 &    2.124 &        - &    3.815 \cr 
  7 &   Pa$_{91}^{233}$ & 0 & 2 & 3 & 4 &    0.061 &    4.720 &  13.833 &  13.394 &       1.238 &    1.238 &    2.262 &    2.226 &        - &    3.804 \cr 
  8 &   U $_{92}^{237}$ & 0 & 2 & 3 & 4 &    0.114 &    4.980 &  13.264 &  12.095 &       0.840 &    0.841 &    1.773 &    1.747 &    3.442 &    2.847 \cr 
  9 &   Np$_{93}^{235}$ & 0 & 2 & 3 & 4 &    0.107 &    5.874 &   8.633 &   7.206 &       0.778 &    0.777 &    1.623 &    1.589 &        - &    2.552 \cr 
 10 &   Np$_{93}^{237}$ & 0 & 2 & 3 & 4 &    0.104 &    5.578 &  10.146 &   8.831 &       0.815 &    0.814 &    1.699 &    1.647 &    3.753 &    2.714 \cr 
 11 &   Np$_{93}^{239}$ & 0 & 2 & 3 & 4 &    0.083 &    5.364 &  11.362 &  10.046 &       0.898 &    0.898 &    1.793 &    1.736 &    4.036 &    2.932 \cr 
 12 &   Pu$_{94}^{239}$ & 0 & 2 & 3 & 4 &    0.115 &    5.883 &   8.964 &   7.583 &       0.784 &    0.783 &    1.659 &    1.599 &    3.386 &    2.656 \cr 
 13 &   Pu$_{94}^{241}$ & 0 & 2 & 3 & 4 &    0.069 &    5.447 &  11.431 &   9.997 &       1.270 &    1.267 &    2.463 &    2.114 &    4.270 &    3.736 \cr 
 14 &   Am$_{95}^{241}$ & 0 & 2 & 2 & 4 &    0.033 &    5.983 &   8.554 &   7.337 &       1.196 &    1.201 &    1.388 &    1.170 &        - &    3.789 \cr 
 15 &   Am$_{95}^{243}$ & 0 & 2 & 3 & 4 &    0.061 &    5.623 &  10.643 &   9.447 &       0.808 &    0.811 &    1.477 &    1.639 &        - &    2.799 \cr 
 16 &   Am$_{95}^{245}$ & 0 & 2 & 3 & 4 &    0.042 &    5.198 &  12.286 &  11.997 &       1.653 &    1.655 &        - &    2.656 &        - &    4.669 \cr 
 17 &   Cm$_{96}^{243}$ & 0 & 2 & 3 & 4 &    0.070 &    6.400 &   7.505 &   5.699 &       1.279 &    1.283 &        - &    2.003 &        - &    3.587 \cr 
 18 &   Cm$_{96}^{245}$ & 0 & 2 & 3 & 4 &    0.093 &    5.908 &  10.041 &   8.278 &       1.242 &    1.241 &    2.437 &    1.956 &    4.075 &    3.477 \cr 
 19 &   Cm$_{96}^{249}$ & 0 & 2 & 3 & 4 &    0.065 &    6.077 &   8.696 &   7.274 &       1.253 &    1.250 &        - &    2.055 &        - &    3.625 \cr 
 20 &   Bk$_{97}^{241}$ & 0 & 2 & 3 & 4 &    0.052 &    7.858 &   2.217 &   0.125 &       0.784 &    0.786 &    1.421 &    1.524 &        - &    2.624 \cr 
 21 &   Bk$_{97}^{247}$ & 0 & 2 & 3 & 4 &    0.042 &    6.597 &   7.079 &   5.161 &       0.935 &    0.930 &    1.390 &    1.703 &        - &    3.008 \cr 
 22 &   Bk$_{97}^{249}$ & 0 & 2 & 3 & 4 &    0.055 &    6.739 &   6.255 &   4.486 &       1.135 &    1.136 &    2.025 &    1.978 &    3.025 &    3.179 \cr 
 23 &   Bk$_{97}^{251}$ & 0 & 2 & 3 & 4 &    0.078 &    6.401 &   7.929 &   6.095 &       0.953 &    0.952 &    1.547 &    1.639 &        - &    2.727 \cr 
 24 &   Cf$_{98}^{247}$ & 0 & 2 & 3 & 4 &    0.075 &    6.945 &   5.978 &   4.061 &       1.258 &    1.259 &    2.296 &    1.900 &        - &    3.339 \cr 
 25 &   Cf$_{98}^{251}$ & 0 & 2 & 3 & 4 &    0.044 &    7.133 &   4.857 &   3.161 &       1.270 &    1.265 &    2.176 &    2.020 &    2.927 &    3.553 \cr 
 26 &   Cf$_{98}^{253}$ & 0 & 2 & 3 & 4 &    0.059 &    6.622 &   6.940 &   5.382 &       1.672 &    1.674 &    2.496 &    2.554 &        - &    4.528 \cr 
\hline
\end{tabular}
\end{table}
\end{center}


\begin{thebibliography}{00}

\bibitem{Del10} D.S. Delion,
{\it Theory of particle and cluster emission}
(Springer-Verlag, Berlin, 2010).
\bibitem{Buc12} D. Bucurescu and N.V. Zamfir,
Phys. Rev. {\bf C 86}, 067306 (2012).
\bibitem{Neu92} R. Neu and F. Hoyler, Phys. Rev. {\bf C 46}, 208 (1992).
\bibitem{Del06} D.S. Delion, S. Peltonen, and J. Suhonen,
Phys. Rev. {\bf C 73}, 014315 (2006).
\bibitem{Del15} D.S. Delion, A. Dumitrescu,
At. Data Nucl. Data Tables {\bf 101}, 1 (2015).
\bibitem{Nil} S.G. Nilsson, Selskab Mat. Fys. Medd. 29 (16) (1955).
\bibitem{Ni12} Dongdong Ni, Zhongzhou Ren, Phys. Rev. {\bf C 86}, 054608 (2012)
\bibitem{War15} D.E. Ward, B. G. Carlsson and S. {\AA}berg, Phys. Rev {\bf C 92}, 014314 (2015).
\bibitem{Sei15} W. M. Seif, M. M. Botros and A. I. Refaie, Phys. Rev. {\bf C92} 044302 (2015).
\bibitem{Lip69} P. O. Lipas and J. Savolainen, Nucl. Phys. A 130, 77 (1969).
\bibitem{Lip76} P. O. Lipas, P. Haapakoski, and T. Honkaranta, Phys. Scr. 13,
339 (1976).
\bibitem{Rad76a} A. A. Raduta and R. M. Dreizler, Nucl. Phys. A 258, 109 (1976).
\bibitem{Rad76b} A. A. Raduta, V. Ceausescu, and R. M. Dreizler, Nucl. Phys. A
272, 11 (1976).
\bibitem{Rad81} A. A. Raduta, V. Ceausescu, A. Gheorghe, and R. M. Dreizler,
Phys. Lett. B 99, 444 (1981).
\bibitem{Rad82} A. A. Raduta, V. Ceausescu, A. Gheorghe, and R. M. Dreizler,
Nucl. Phys. A 381, 253 (1982).
\bibitem{Rad12} A. A. Raduta, R. Budaca, and Amand Faessler, Ann. Phys. (NY)
327, 671 (2012).
\bibitem{Rad14} A.A. Raduta, {\it Nuclear Structure with Cohent States}
(Springer International Publishing, Switzerland, 2015).
\bibitem{Del93} A. A. Raduta, D. S. Delion and N. Lo Iudice, Nucl. Phys. {\bf A551} (1993).
\bibitem{Ber77} G. Bertsch, J. Borysowicz, H. McManus, and W.G. Love,
Nucl. Phys. A {\bf 284}, 399 (1977).
\bibitem{Sat79} G.R. Satchler and W.G. Love, Phys. Rep. {\bf 55}, 183 (1979).
\bibitem{Car92} F. C\^arstoiu and R.J. Lombard,
Ann. Phys. (N.Y.) {\bf 217}, 279 (1992).
\bibitem{Pel08} S. Peltonen, D.S. Delion, and J. Suhonen,
Phys. Rev.  {\bf C 78}, 034608 (2008).
\bibitem{Del15a} D. S. Delion, A. Dumitrescu,
Phys. Rev. {\bf C 92}, 021303(R) (2015).
\bibitem{Bnl}\textit{Evaluated Nuclear Structure Data Files} at Brookhaven National Laboratory,
www.nndc.bnl.gov/ensdf/.
\bibitem{Del09} D. S. Delion, Phys. Rev. {\bf C} 80, 024310 (2009).


\end{thebibliography}
\end{document}